\documentclass[aip,pop,reprint]{revtex4-1}
\usepackage{graphicx}
\usepackage{amsmath}
\usepackage{mathrsfs}
\usepackage{array}
\usepackage[breaklinks=true,colorlinks=true,linkcolor=blue,urlcolor=blue,citecolor=blue]{hyperref}
\graphicspath{{./figures/}}

\begin{document}

\title{Measurement of wave-particle interaction and metastable lifetime using laser-induced fluorescence} 
\author{F. Chu}
\email[]{feng-chu@uiowa.edu}
\affiliation{Department of Physics and Astronomy, University of Iowa, Iowa City, Iowa 52242, USA}
\author{R. Hood}
\affiliation{Department of Physics and Astronomy, University of Iowa, Iowa City, Iowa 52242, USA}
\author{F. Skiff}
\affiliation{Department of Physics and Astronomy, University of Iowa, Iowa City, Iowa 52242, USA}
\date{\today}

\begin{abstract}
\rightskip.5in
Extensive information, such as ion temperature and wave-particle interactions, can be obtained by direct measurement of ion distribution functions using laser-induced fluorescence (LIF). This nonintrusive plasma diagnostic provides an important window into the ion motions in phase-space. Previous simulation results suggest that LIF measurements, which are performed on metastable ions produced directly from neutral gas particles and also from ions in other electronic states, place restrictions on the metastable lifetime. In the case where metastable population is produced from direct ionization of neutral atoms, the velocity distribution measured using LIF will only faithfully represent processes which act on the ion dynamics in a time shorter than the metastable lifetime. In this paper, the metastable lifetime effects are explored experimentally for the first time using wave-particle interaction in an Argon multidipole plasma. We demonstrate that this systematic effect can be corrected using the theory addressed in the paper based on the metastable lifetime and relative fraction of metastables produced from pre-existing ions.
\end{abstract}

\maketitle

\section{Introduction}
\label{sec:intro}

Laser-induced fluorescence (LIF) is a nonintrusive plasma diagnostic technique used to probe the dynamics of ion motions in phase-space. As long as the effect of photon momentum on ion orbits is negligible \cite{hollmann_measurement_1999}, LIF measurements do not disturb the plasma, enabling the accurate study of velocity-space diffusion \cite{bowles_direct_1992}, ion acoustic wave reflection \cite{berumen_analysis_2018}, ion heating \cite{anderegg_ion_1986, mcchesney_observation_1987}, and other related phenomena in plasmas \cite{skiff_direct_1987, severn_experimental_2003}. For example, LIF is extensively employed to determine the concentrations of chemical species in industrial etch reactors where nonperturbative measurements are required \cite{booth_spatially_1989, mcwilliams_laser-induced_2007}. This laser diagnostic is also frequently adopted to examine plume plasma properties of Hall effect thrusters to study the physical processes that control their operation \cite{w._a._hargus_laser-induced_2001, mazouffre_time-resolved_2009}. 

In order to induce fluorescence, a laser is tuned to a specific wavelength to resonate with metastable ions produced from direct ionization of neutral atoms as well as ions in other electronic states. Metastables produced from existing ions have a history as “typical” ions. However, the newly produced metastables from neutral atoms are representative of the neutral velocity distribution and only become “typical” over time through ion-ion coulomb collisions. Due to the finite metastable lifetime, metastable ions can have a different velocity distribution from the ground-state ions (the majority ion population in many laboratory plasmas), causing errors in LIF detections of any observable quantities that are derived from the ion velocity distribution function, such as ion temperature and amplitude of electrostatic waves. Here rises a fundamental issue in LIF measurements: when can Doppler-resolved LIF on metastable ions be used to infer the velocity distribution of ground-state ions? This important matter is the basis of LIF that concerns anyone who uses LIF as a plasma diagnostic.

Numerical simulations based on our newly developed Lagrangian model for LIF \cite{claire_nonlinear_2001, chu_determining_2017, chu_lagrangian_2018} show that under circumstances where the metastable ion population is produced from direct ionization of neutral atoms, the velocity distribution measured using LIF will only faithfully represent processes which act on the ion dynamics in a time shorter than the metastable lifetime. For instance, the LIF measured ion temperature on these metastables is only accurate when they live longer than the ion-ion collision mean free time. For the purposes of wave detection \cite{sarfaty_direct_1996, mcwilliams_experimental_1986}, the wave period has to be shorter than the metastable lifetime for the perturbed distribution $f_1(v,t)$ measured on this metastable population to be correct. However, the LIF measurements made on the metastables produced from pre-existing ions are not affected by their lifetime.

A good understanding of the behavior associated with each metastable population helps us avoid the systematic errors caused by the finite metastable lifetime. When these errors become inevitable, correction of the LIF measurements requires knowledge on the metastable lifetime and fraction of metastables produced from pre-existing ions as opposed to directly from neutral atoms.

In this paper, we present the first experiment on determining the metastable lifetime as well as the relative fraction of metastables produced from pre-existing ions in an Argon multidipole plasma. The technique relies on measuring the ionic wave response. The first numerical study of the hole-burning effect on the metastable distribution perturbed by an electrostatic wave is also reported. The paper is organized as follows: Sec.~\ref{sec:theosim} presents a simple theory for the metastable lifetime effects and the comparison between the theory and the simulation results, Sec.~\ref{sec:exp} gives a description of the experimental setup, Sec.~\ref{sec:results} presents the experimental results, and Sec.~\ref{sec:summary} provides a summary.

\section{Theory and Simulation}
\label{sec:theosim}

As mentioned earlier, metastable ions are produced from direct ionization of neutral atoms as well as ions in other electronic states. Assuming the plasma is weakly collisional and spatially uniform, in the presence of an electrostatic wave the velocity distribution functions of metastable ions produced from neutrals and pre-existing ions, $p$ and $q$, can be obtained by solving the modified Vlasov equation for each population
\begin{gather}
\label{eq:f1Ve}
\frac{\partial p}{\partial t}+v\frac{\partial p}{\partial x}+\frac{Ee}{m_\textup{i}}\frac{\partial p}{\partial v}=S_\textup{n}f_\textup{n}-\xi p, \\[2ex]
\label{eq:f2Ve}
\frac{\partial q}{\partial t}+v\frac{\partial q}{\partial x}+\frac{Ee}{m_\textup{i}}\frac{\partial q}{\partial v}=S_\textup{i}f_\textup{i}-\xi q,
\end{gather}
where $S_\textup{n}$ and $S_\textup{i}$ are the birth rates of metastables produced from neutrals and pre-existing ions, $E$ is the amplitude of the electrostatic wave, $\omega$ is the wave angular frequency, and $m_{\textup{i}}$ is the ion mass. The velocity distributions of the neutral atoms and ground-state ions are denoted by $f_\textup{n}$ and $f_\textup{i}$, respectively. The metastable ions produced from pre-existing ions have the same velocity distribution as ground-state ions, i.e., $q=f_\textup{i}$. According to Ref.~\onlinecite{chu_lagrangian_2018}, the metastable lifetime $\tau$ is limited by metastable quench rate $r$, electron-collisional excitation rate $u$, and optical pumping rate $W$ (from the metastable state to the upper state)
\begin{equation}
\label{eq:lifetime}
1/\tau=\xi=r+\gamma \left ( u+W \right ),
\end{equation} 
where $\gamma =1-A_\textup{21}/A_\textup{T}$, $A_\textup{21}$ is the Einstein coefficient of spontaneous decay from the upper state to the metastable state, and $A_\textup{T}$ is the total spontaneous decay rate of the upper state.


Laboratory plasmas are often in the regime where the ion sound speed is much larger than the ion thermal speed. If we further assume that the neutrals and ions have the same temperature (both neutrals and ions have the same zeroth-order distribution $f_0$) and ignore the wave vector $\textbf{k}$, one can compute the first-order perturbation of $p$ and $q$ as
\begin{gather}
\label{eq:p}
p=- \frac{iEe}{m_{\textup{i}}(\omega+i\xi)} \cdot \frac{\partial f_0}{\partial v}, \\[2ex]
\label{eq:q}
q=- \frac{iEe}{m_{\textup{i}}\omega} \cdot \frac{\partial f_0}{\partial v}.
\end{gather}
Since LIF directly measures the distribution of the metastable ions, therefore, the first-order perturbation of the LIF measured ion distribution in the presence of an electrostatic wave can be written as
\begin{equation}
\label{eq:f1LIF}
f_{\textup{1-LIF}}=- \frac{iEe}{m_{\textup{i}}} \cdot \frac{\partial f_0}{\partial v}\left ( \frac{n_{\textup{meta-i}}}{\omega}+ \frac{n_{\textup{meta-n}}}{\omega+i\xi } \right ),
\end{equation}
where the densities of the metastable ions produced from neutrals and pre-existing ions are denoted by $n_{\textup{meta-n}}$ and $n_{\textup{meta-i}}$, respectively. By following the same procedure, the second-order perturbation of the LIF measured ion distribution is
\begin{equation}
\label{eq:f2LIF}
f_{\textup{2-LIF}}=- \left (  \frac{Ee}{m_{\textup{i}}} \right )^2 \cdot \frac{\partial^2 f_0}{\partial v^2}\left [ \frac{n_{\textup{meta-i}}}{\omega^2}+ \frac{n_{\textup{meta-n}}}{(\omega+i\xi)^2 } \right ].
\end{equation}

In a general case where the neutral atoms and ions are not in thermal equilibrium, it is difficult to find the exact solutions of $f_{\textup{1-LIF}}$ and $f_{\textup{2-LIF}}$. However, their numerical solutions can still be calculated using the Lagrangian model for LIF as described in Sec.~\ref{subsec:sim}.

\subsection{Metastable Lifetime Effects}
\label{subsec:lifetime}

The perturbed ion distributions measured using LIF in Eqs.~(\ref{eq:f1LIF})--(\ref{eq:f2LIF}) reveal different behaviors of the two metastable populations. In the limit when the metastable lifetime is much longer than the wave period ($\tau \gg 1/\omega$), the LIF measured first-order perturbation $f_{\textup{1-LIF}}$ is proportional to the total metastable density $n_{\textup{meta-i}}+n_{\textup{meta-n}}$
\begin{equation}
\label{eq:limit1}
\lim_{\omega \gg \xi}  f_{\textup{1-LIF}} = - \frac{iEe}{m_{\textup{i}}\omega} \cdot \frac{\partial f_0}{\partial v}\left ( n_{\textup{meta-i}}+n_{\textup{meta-n}} \right ).
\end{equation}
However, in the other limit when the metastable lifetime is short ($\tau \ll 1/\omega$), $f_{\textup{1-LIF}}$ is proportional to $n_{\textup{meta-i}}$ with no contribution from $n_{\textup{meta-n}}$ at all
\begin{equation}
\label{eq:limit2}
\lim_{\omega \ll \xi}  f_{\textup{1-LIF}} = - \frac{iEe}{m_{\textup{i}}\omega} \cdot \frac{\partial f_0}{\partial v} n_{\textup{meta-i}}.
\end{equation}

The difference between these two metastable populations results from their distinctive histories. The lifetime of the metastables produced from neutrals sets the time scale they experience the wave field. With a history of being neutral particles, this population cannot react to the electric field until they become ions. If the lifetime is shorter than the wave period, these metastables will not live long enough to interact with the wave, resulting in a reduction in the measured $f_1$. On the other hand, as metastables produced from pre-existing ions have already fully interacted with the wave field before becoming metastables, the perturbed distribution measured using LIF is independent of their lifetime. This analysis can also be applied to $f_{\textup{2-LIF}}$ in Eq. (\ref{eq:f2LIF}).

\begin{figure}
\begin{center}
\includegraphics[width=3.37in]{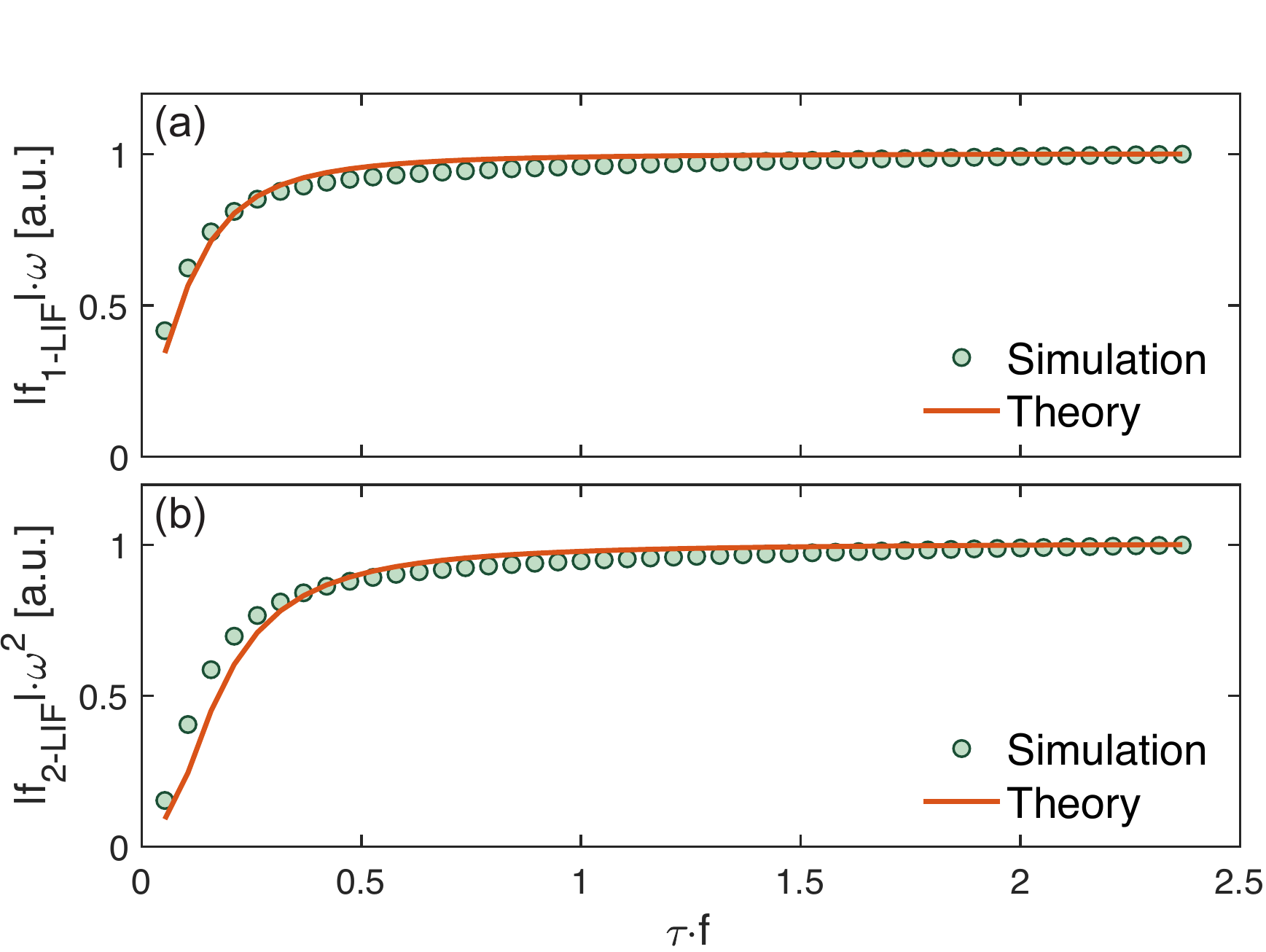}
\caption{Comparison between the Lagrangian model and the theory for $f_{\textup{1-LIF}}$ and $f_{\textup{2-LIF}}$ at various metastable lifetimes is shown in (a) and (b), respectively. The metastable lifetime $\tau$ is normalized by the wave frequency $f=\omega/2\pi$. The relative fraction of metastables produced from pre-existing ions $n_{\textup{meta-i}}/(n_{\textup{meta-i}}+n_{\textup{meta-n}})=14.3$ \%. Reproduced with permission from Phys. Rev. Lett. \textbf{122}, 075001 (2019). Copyright 2019 American Physical Society \cite{chu_determining_2019}.}
\label{fig:simulation-theory}
\end{center}
\end{figure}

According to Eqs. (\ref{eq:f1LIF})--(\ref{eq:f2LIF}), the LIF measured amplitude of the electrostatic wave can be calculated as
\begin{equation}
\label{eq:ELIF}
E_{\textup{LIF}}= \left | \left (f_{\textup{2-LIF}}\cdot \frac{\partial f_0}{\partial v} \right ) \bigg/ \left ( f_{\textup{1-LIF}}\cdot \frac{\partial^2 f_0}{\partial v^2} \right ) \right | \cdot \frac{m_{\textup{i}}\omega }{e}.
\end{equation}
As expected, $E_{\textup{LIF}}$ is as well affected by the metastable lifetime effects. By comparing $E_{\textup{LIF}}$ with the measurement of the same wave electric field using a different method not relying on metastable ions, such as an electric field probe, the metastable lifetime effects can therefore be observed experimentally.

\subsection{Simulation}
\label{subsec:sim}

\begin{figure*}
\begin{center}
\includegraphics[width=6.69in]{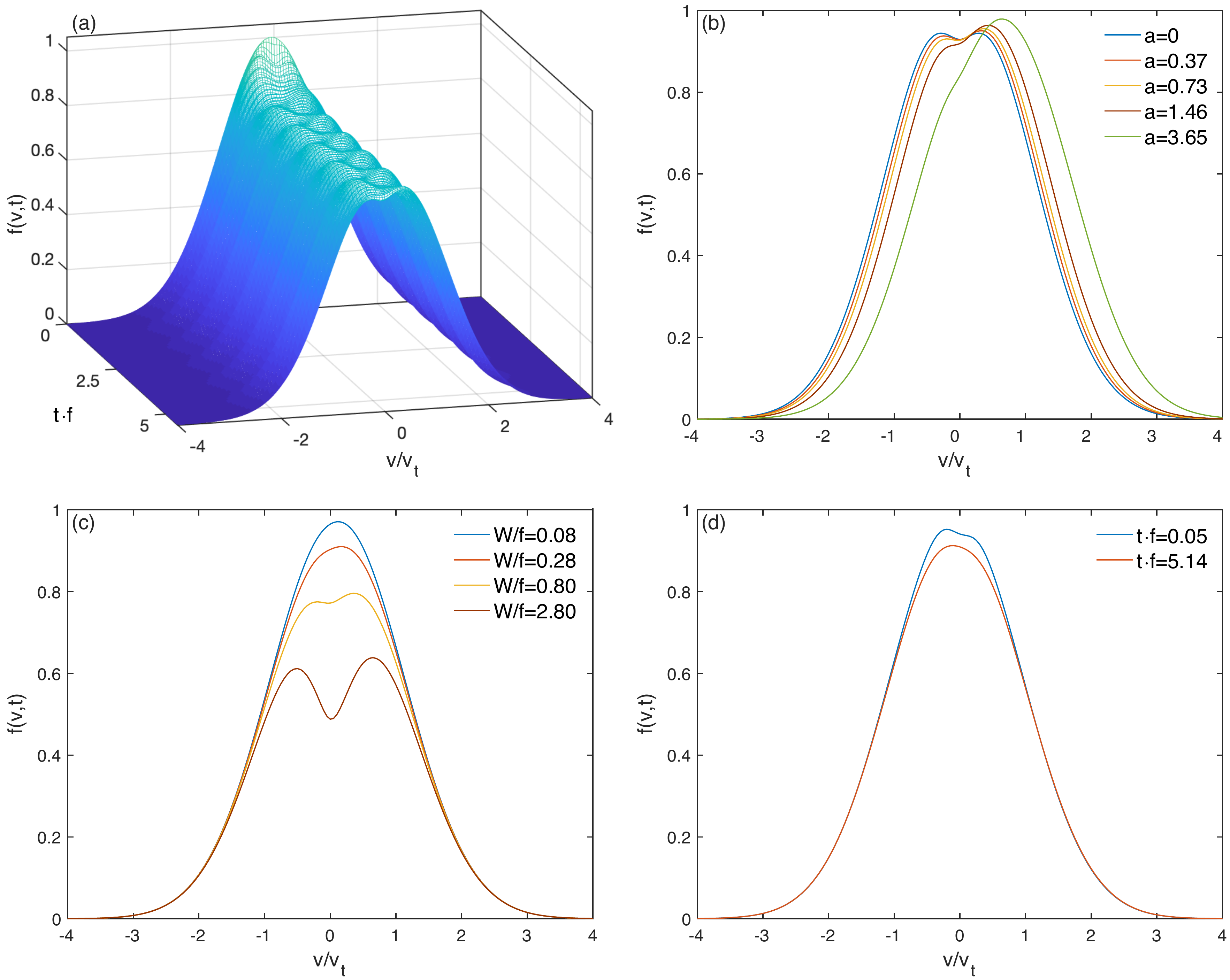}
\caption{Simulation results of the hole-burning effect using the Lagrangian model for LIF. The frequency of the laser is in resonance with ions at $v=0$. The fraction of metastables produced from pre-existing ions $n_{\textup{meta-i}}/(n_{\textup{meta-i}}+n_{\textup{meta-n}})=50$ \%. (a) Time evolution of the metastable distribution $f(v,t)$ perturbed by electrostatic waves during optical pumping. The velocity $v$ and time $t$ are normalized by the thermal speed $v_\textup{t}$ and wave frequency $f$, respectively. The optical pumping rate $W/f$ at the peak of $f(v,t)$ is 0.80. The amplitude of the electrostatic wave $a=Ee/m_{\textup{i}}v_\textup{t}f=0.73$. (b) Metastable distribution perturbed by electrostatic waves with different wave amplitudes. The normalized time $tf$ is selected at 5.46 when the peak of $f(v,t)$ is in the maximum displacement from $v=0$. The optical pumping rate $W/f=0.80$. (c) Metastable distribution under different optical pumping rates. The normalized time and amplitude of the wave are $tf=5.46$ and $a=0.73$, respectively. (d) Metastable distribution at different time $t$. The optical pumping rate and amplitude of the wave are $W/f=0.28$ and $a=0.73$, respectively.}
\label{fig:holeburning}
\end{center}
\end{figure*}


To test the theory for $f_{\textup{1-LIF}}$ and $f_{\textup{2-LIF}}$ in Eqs.~(\ref{eq:f1LIF})--(\ref{eq:f2LIF}), a numerical simulation is performed based on a Lagrangian approach for LIF, developed specifically to model the LIF measurements in laboratory plasmas. A detailed description of the model and its application are presented in Ref.~\onlinecite{chu_determining_2017} and \onlinecite{chu_lagrangian_2018}, so here we only give a brief overview.

The Lagrangian approach is an interpretation of LIF where one follows each individual ion orbit as it moves through space and time. Unlike the traditional Eulerian model which extends the rate equations to a system of coupled kinetic equations, the Lagrangian model separates the classical dynamics of the ions from the quantum mechanics of the electronic states. This separation provides a large computational advantage, as it reduces the coupled partial differential equations in the Eulerian model to ordinary differential equations. In addition, since the Lagrangian model does not impose constraints on the ion orbits, it can be applied to systems with complicated ion dynamics. The simulation results for $f_{\textup{1-LIF}}$ and $f_{\textup{2-LIF}}$ demonstrate a good agreement with the theoretical predictions, as shown in Fig. \ref{fig:simulation-theory}.

\subsection{Discussion}

\begin{figure*}
\begin{center}
\includegraphics[width=6.69in]{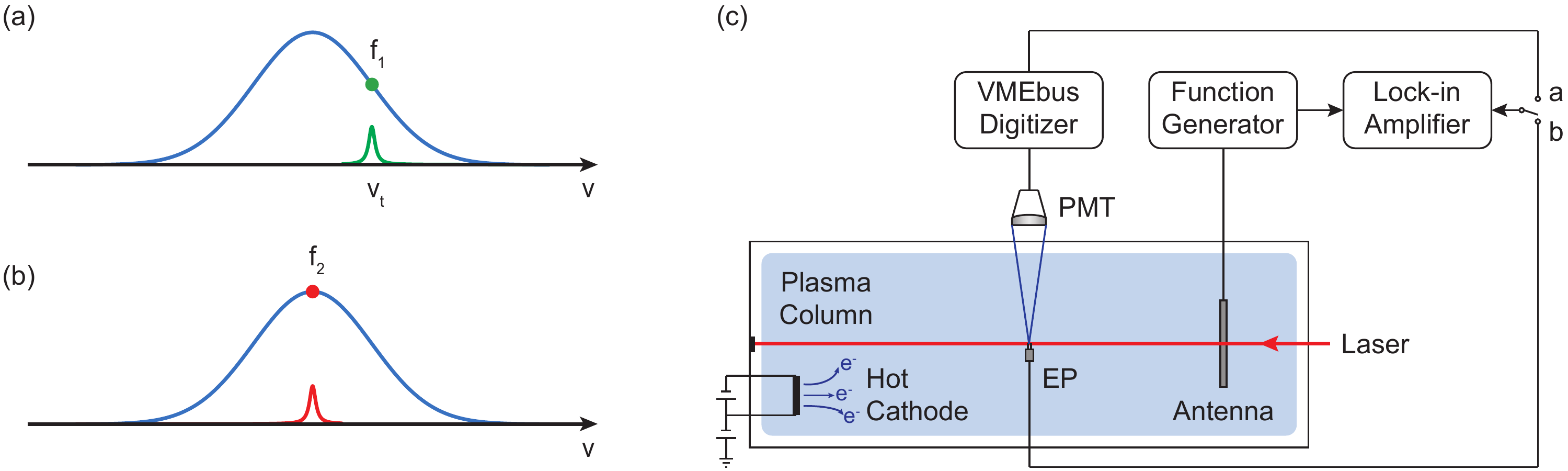}
\caption{Frequencies of the laser selected to resolve $f_1$ and $f_2$ in velocity-space in the experiment are shown in (a) and (b), respectively; $f_1$ is measured at thermal velocity $v_\textup{t}$ and $f_2$ at the peak of $f_0$. (c) Schematic of the experimental setup. The LIF diagnostic equipment (not drawn to scale) includes a double-mesh antenna of 65 $\%$ open area, a 16-channel photomultiplier tube (PMT), a Versa Module Europa bus (VMEbus) board, and a lock-in amplifier. A disc-shaped Langmuir probe with a diameter of 0.65 cm is placed in the bulk plasma to measure the electron density and temperature. A differential sinusoidal signal with $V_\textup{p}=\pm2$ V is applied on the double-mesh antenna that is 5 cm away from the LIF viewing volume to excite ion acoustic waves in the plasma. The wave electric field is measured using both LIF and a double-tip electric field probe (EP).}
\label{fig:exp}
\end{center}
\end{figure*}

In LIF measurements, the fluorescence signal is produced from allowed transitions of plasma ions that are optically pumped to excited states. This optical pumping process depletes the metastable ion population through ``burning a hole'' in its distribution function \cite{bennett_hole_1962}. Since the LIF measured perturbed ion distribution $f_{\textup{1-LIF}}$ and $f_{\textup{2-LIF}}$ are used to construct $E_{\textup{LIF}}$ in Eq.~(\ref{eq:ELIF}), it is necessary to understand how the metastable distribution is affected by optical pumping in the presence of an electrostatic wave. 


We simulate the hole-burning effect using the Lagrangian model for LIF and the results are presented in Fig.~\ref{fig:holeburning}. To our knowledge, this is the first numerical study of the optical pumping process involving an electrostatic wave. The frequency of the laser selected in the simulation is in resonance with ions at $v=0$. The time evolution of the metastable distribution perturbed by an electrostatic wave is shown in Fig.~\ref{fig:holeburning}(a). The metastable ions initially have a Maxwellian distribution, and then a hole starts to form at $v=0$ as optical pumping pumps those ions to the excited state. The size of the hole keeps growing until optical pumping reaches an equilibrium with velocity-space diffusion and the appearance of new metastables in the laser beam. The metastable distribution perturbed by electrostatic waves with different wave amplitudes is shown in Fig.~\ref{fig:holeburning}(b). This plot demonstrates that the size of the hole is not only affected by optical pumping but also the wave amplitude. Figure~\ref{fig:holeburning}(c) shows the metastable distribution under different optical pumping rates. When the metastable distribution is not perturbed by the electrostatic wave, optical pumping always burns a hole in the distribution function. However, when affected by wave electric fields, the resulting shape of the distribution function in the optical pumped region can be either concave or convex. The metastable distribution at different times is plotted in Fig.~\ref{fig:holeburning}(d). It shows that optical pumping burns a hole in the metastable distribution at the beginning, however, the hole is averaged out later as the ions are sloshed back and forth in velocity-space by the wave field.

Since the second derivative of a function directly reflects its concavity, the LIF measured second-order perturbation $f_{\textup{2-LIF}}$ is more sensitive to the shape of the metastable distribution function than $f_{\textup{1-LIF}}$. 

\section{Experimental Setup}
\label{sec:exp}

The experiments of measuring the wave-particle interaction and metastable lifetime are performed in an Argon multidipole plasma confined in a cylindrical chamber of 73 cm length and 49 cm diameter \cite{hood_ion_2016}. The plasma is produced through impact ionization by primary electrons emitted from a hot cathode biased at $-70$ V with respect to the chamber walls, resulting in an emission current of 56 mA. The cathode is made of a sintered lanthanum-hexaboride (LaB6) ring segment which is heated resistively. The multidipole magnetic field is provided by an electrically grounded magnet cage consisting of 16 rows of magnets with alternating poles covering all inside walls of the chamber. The magnetic field strength is about 1000 G close to the magnets and quickly diminishes to less than 2 G in the measurement region. Neutral pressure is regulated by a mass flow controller and measured using an ionization gauge.

%

The LIF scheme used in the experiment is accomplished by a single mode tunable Rhodamine 6G dye laser (Sirah Matisse-DS). The fluorescence signal is observed through a 25 cm diameter window, filtered using a narrow bandwidth interference filter, and imaged onto a 16-channel photomultiplier tube (PMT). To induce fluorescence, in the rest frame of an ion, the laser is tuned at 611.662 nm to excite electrons in the metastable state $\textup{3d }^2\textup{G}_{9/2}$ to the upper state $\textup{4p }^2\textup{F}_{7/2}^{o}$. Fluorescence photons are emitted at 461.086 nm when those electrons decay to the $\textup{4s }^2\textup{D}_{5/2}$ state with a large branching ratio of $66.5$ \% \cite{severn_argon_1998, mattingly_measurement_2013}. In principle the LIF measured electric field $E_{\textup{LIF}}$ in Eq. (\ref{eq:ELIF}) can be obtained by sampling $f_1$ and $f_2$ at almost any point in velocity-space. However, to maximize the fluorescence signal the LIF measurements of $f_1$ and $f_2$ are made at $v_\textup{t}$ (ion thermal speed) and the peak of $f_0$ respectively, as illustrated in Figs.~\ref{fig:exp}(a)--\ref{fig:exp}(b).

\begin{figure}[b]
\begin{center}
\includegraphics[width=3.37in]{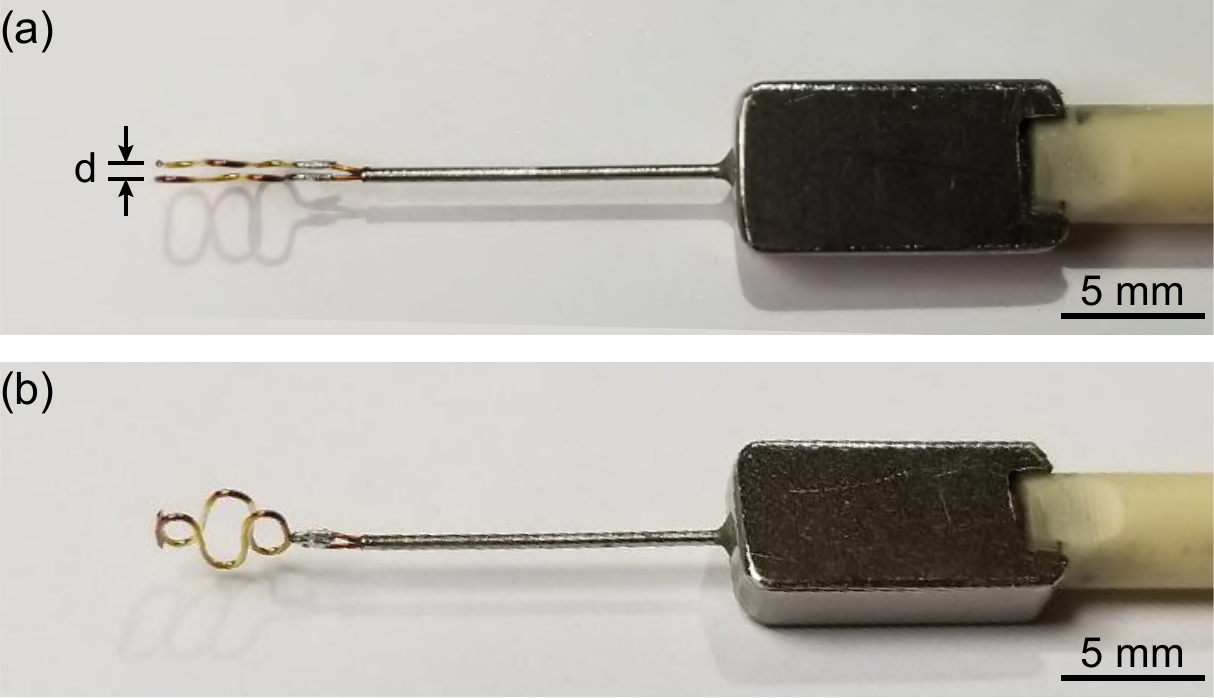}
\caption{Photographs of the double-tip electric field probe. The instrumentation amplifier AD8421 is inclosed in the metal box located on the right side of the photographs. The length from the probe tips to the instrumentation amplifier is only about 15 mm. Top view of the electric field probe is shown in (a). The separation between the tips $d \approx 0.5 \text{ } \textup{mm}$. Side view in (b) shows the shape of the probe tips, which are designed with an S shape to prevent shadowing in the direction of the wave propagation.}
\label{fig:probe}
\end{center}
\end{figure}


The experimental setup is depicted in Fig.~\ref{fig:exp}(c). A differential sinusoidal signal with $V_\textup{p}=\pm2$ V is applied on the double-mesh antenna to excite ion acoustic waves in the plasma. The same signal is also sent to the lock-in amplifier as a reference. To measure the response of metastable ions to the wave electric field at various frequencies, the wave is scanned from 1 kHz to 45 kHz with an increment of 1 kHz. At each wave frequency $\omega$, $f_1$ and $f_2$ are resolved using the lock-in amplifier by locking the frequency at $\omega$ and $2\omega$ respectively. The LIF measured electric field $E_{\textup{LIF}}$ can therefore be calculated using Eq.~(\ref{eq:ELIF}). The same electric field is also evaluated using a double-tip electric field probe to compare with the LIF measurements.

The electric field probe shown in Fig.~\ref{fig:probe}(a)--\ref{fig:probe}(b) is made with a low noise, high speed instrumentation amplifier AD8421 ideally suited for a broad spectrum of signal conditioning and data acquisition applications. The amplifier features extremely high common-mode rejection ratio, allowing it to extract low level differential voltage signals in the presence of high frequency common-mode noise over a wide frequency range. The probe tips are designed with an S shape to prevent from shadowing each other in the direction of the wave propagation. As plasmas tend to have a large impedance, even small capacitance from the wires connecting the probe tips and the instrumentation amplifier can significantly reduce the bandwidth of the probe. Therefore, to improve the probe's performance in the high frequency range, the instrumentation amplifier is placed only 15 mm away from the tips. The electric field probe measures the differential voltage between two points in the direction of the wave propagation and gives the electric field $E_{\textup{probe}}=V_{\textup{out}}/dG$, where $V_\textup{out}$ is the output voltage of the probe and $G=100$ is the gain of the instrumentation amplifier. The separation $d$ between the probe tips is about half a millimeter, providing an excellent spatial resolution in the electric field measurements. Because of the large plasma impedance comparable to the input impedance of the instrumentation amplifier and difficulties in precisely measuring the tip separation, the probe measured electric field $E_{\textup{probe}}$ also needs to be multiplied by a correction factor $\alpha$ to compensate for those errors.


\section{Experimental Results}
\label{sec:results}

By scanning the laser wavelength from 611.65785 nm to 611.66585 nm, it is found that the ions have a Maxwellian velocity distribution along the direction of the laser beam in the bulk plasma. Based on the halfwidth of the distribution function, the ion temperature is given by $T_\textup{i}=0.03 \pm 0.01$ eV, suggesting that both ions and neutrals are close to the room temperature 0.025 eV. A 0.65 cm diameter disc-shaped Langmuir probe is used to measure electron density $n_\textup{e}$, electron temperature $T_\textup{e}$, and plasma potential $V_\textup{p}$ in the experiment. When the neutral pressure is regulated at $P=0.058 \pm 0.006$ mTorr, the typical plasma parameters in the bulk are, $n_\textup{e}=2.1 \times 10^9$ $\textup{cm}^{-3}$, $T_\textup{e}=2.9$ eV, and $V_\textup{p}=-4.3$ V. The ion sound speed is estimated as $C_\textup{s} \approx \sqrt{T_\textup{e}/m_\textup{i}}=2.6 \times 10^5$ $\textup{cm/s}$, which is much larger than the ion thermal speed $v_\textup{t} = \sqrt{T_\textup{i}/m_\textup{i}} \approx 2.7 \times 10^4$ $\textup{cm/s}$. Therefore both assumptions made in the derivation of Eqs. (\ref{eq:f1LIF})--(\ref{eq:f2LIF}) are satisfied.


\subsection{Comparison between $\mathbf{E_{\textup{LIF}}}$ and $\mathbf{E_{\textup{probe}}}$}
\label{subsec:comparison}

\begin{figure}
\begin{center}
\includegraphics[width=3.37in]{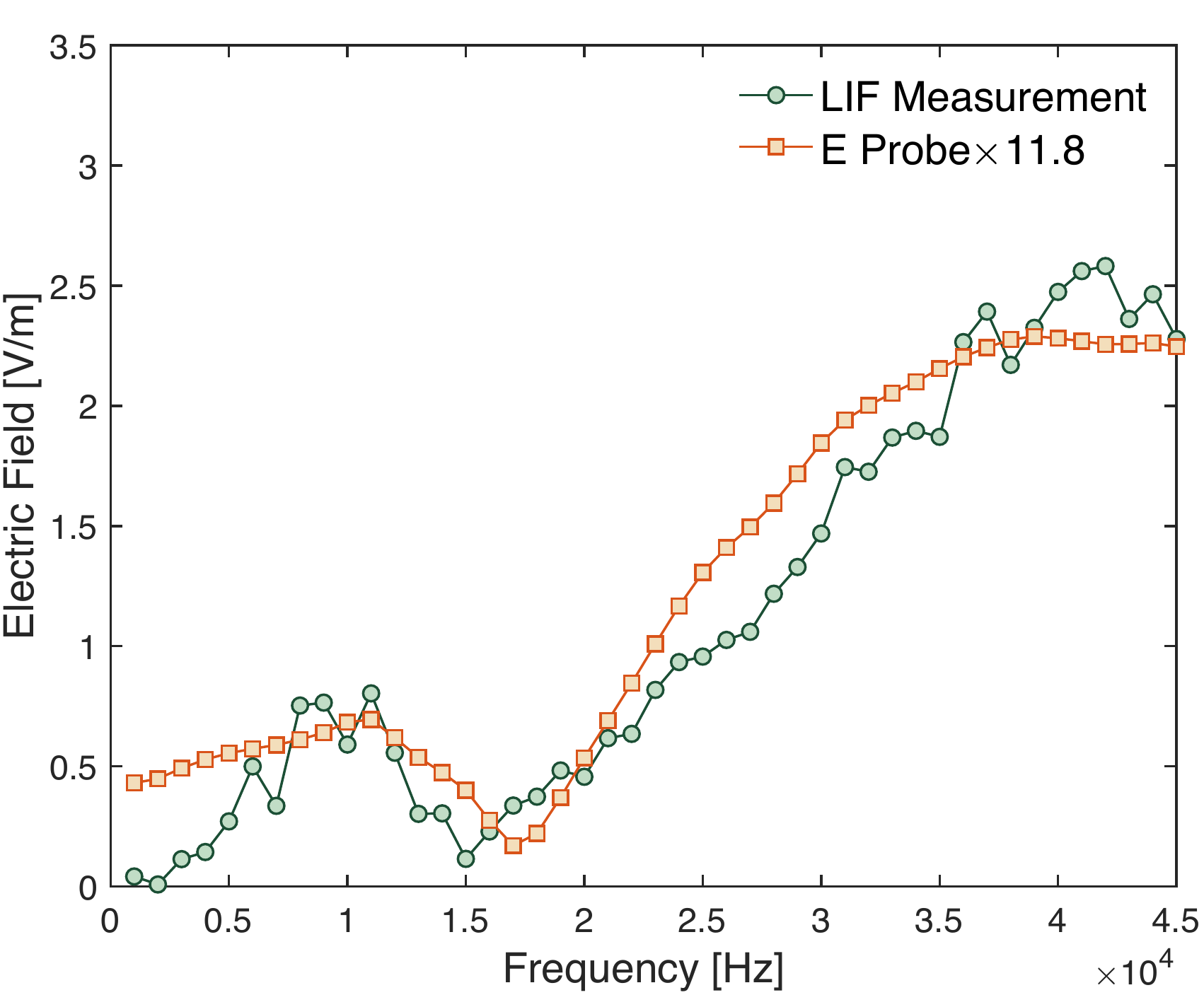}
\caption{Comparison of the ion acoustic wave electric field measured using LIF and the electric field probe at different frequencies. The probe measurement $E_{\textup{probe}}$ is multiplied by a correction factor $\alpha=11.8$ to scale with the LIF measurement $E_{\textup{LIF}}$. Reproduced with permission from Phys. Rev. Lett. \textbf{122}, 075001 (2019). Copyright 2019 American Physical Society \cite{chu_determining_2019}.}
\label{fig:E}
\end{center}
\end{figure}

Figure~\ref{fig:E} shows the electric field of the ion acoustic wave measured using LIF, $E_{\textup{LIF}}$, as long as the probe measurement $E_{\textup{probe}}$ which is multiplied by a correction factor $\alpha=11.8$ to scale. The fast electrons accelerated by the antenna can be picked up by the probe as well, introducing errors to the electric field measurement especially when the field is small. For this reason, the dips on the two curves are slightly displaced from each other around 16 kHz. The electric field measured using the two methods are in good agreement above 10 kHz, however, the LIF measurement is systematically smaller than the probe measurement below 10 kHz due to the metastable lifetime effects discussed in Sec.~\ref{subsec:lifetime}.

\begin{figure}[!h]
\begin{center}
\includegraphics[width=3.37in]{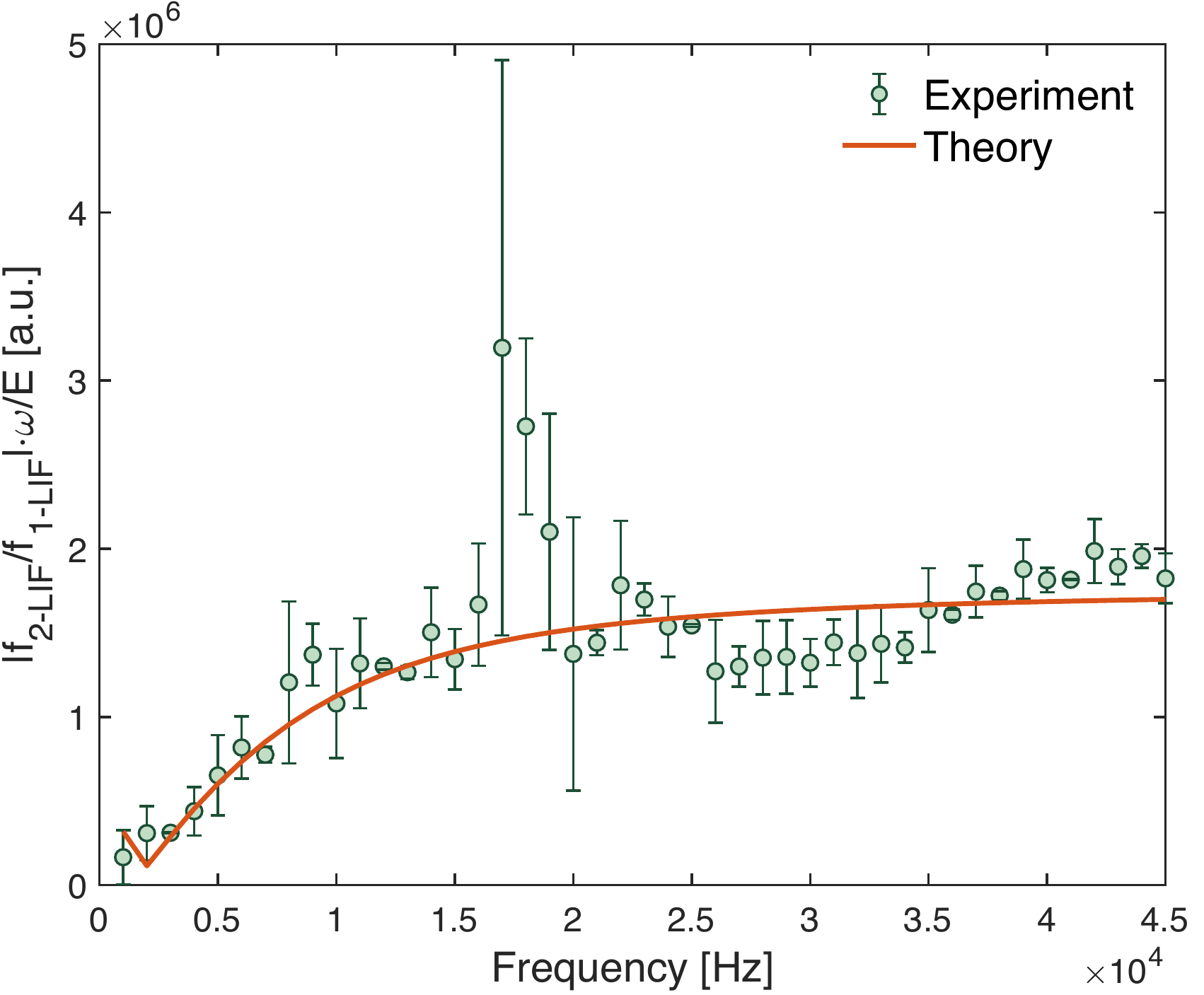}
\caption{LIF measurement of the wave electric field $E_{\textup{LIF}}$ normalized by the probe measurement $E_{\textup{probe}}$. Error bars represent one-standard-deviation uncertainties. The theoretical prediction is also plotted here for comparison. Reproduced with permission from Phys. Rev. Lett. \textbf{122}, 075001 (2019). Copyright 2019 American Physical Society \cite{chu_determining_2019}.}
\label{fig:f2-f1}
\end{center}
\end{figure}

\subsection{Measurement of the Metastable Lifetime}

In the experiment, the efficiency of the antenna in launching the ion acoustic wave varies with the frequency. To compare the experiment with the theory in Eq.~(\ref{eq:ELIF}), the LIF measured electric field $E_{\textup{LIF}}$ needs to be normalized by the probe measurement $E_{\textup{probe}}$. The result of this procedure, shown in Fig.~\ref{fig:f2-f1}, is the key experimental result of this paper. The peak at 16 kHz is caused by the misalignment of the dips in Fig.~\ref{fig:E}. The normalized LIF measurement $E_{\textup{LIF}}/E_{\textup{probe}}$, instead of remaining as a constant, rolls off below 10 kHz due to the finite metastable lifetime.

By fitting the ratio $E_{\textup{LIF}}/E_{\textup{probe}}$ with the theoretical prediction, the metastable lifetime as long as the fraction of metastables produced from pre-existing ions can be determined. Because of the coulomb collisional drag effect (not present for $v$ at the peak of $f_0$), metastable ions with $v \sim v_\textup{t}$ systematically spend less time in resonance with the laser, resulting in a smaller optical pumping rate and a longer lifetime. From the best fit, the inverse metastable lifetime for ions at the peak of $f_0$ is determined as $\xi=(5.63 \pm 0.35) \times 10^4$ $\textup{s}^{-1}$, which gives the metastable lifetime $\tau=17.8 \pm 1.1$ $\mu \textup{s}$. For metastable ions at thermal velocity, these two parameters are $\xi=(4.09 \pm 0.23) \times 10^4$  $\textup{s}^{-1}$ and $\tau=24.4 \pm 1.4$ $\mu \textup{s}$. The sum of the quench rate and collisional excitation rate $r+\gamma  u$ is estimated as $1 \times 10^4$ $\textup{s}^{-1}$, which is at least three times smaller than the optical pumping rate $W$, making the latter the dominant factor in controlling the metastable lifetime in the experiment ($\gamma \approx 0.83$ for 611.662 nm line). Finally, the relative fraction of the metastables produced from pre-existing ions is estimated as $n_{\textup{meta-i}}/(n_{\textup{meta-i}}+n_{\textup{meta-n}})=4 \pm 2$ \%. This result suggests that the metastable ions are mainly produced by direct ionization of neutrals in this Argon multidipole plasma \cite{goeckner_laserinduced_1991}.


To validate the above results, the theoretical quench rate and collisional excitation rate are computed to compare with the experimental values. Based on collision theory, the quench rate of metastable ions is given by \cite{moore_physical_1963}
\begin{equation}
\label{eq:quench}
r=n_{\textup{n}} \sigma \sqrt{\frac{8T_{\textup{n}}}{\pi m_{\textup{n}}}},
\end{equation}
where $n_{\textup{n}}$ is the neutral density, $\sigma$ is the quench cross section, $T_{\textup{n}}$ is the neutral temperature in energy units, and $m_{\textup{n}}$ is the mass of the neutral particles. According to Ref.~\onlinecite{skiff_ion_2001}, $\sigma=2.7 \times 10^{-14}$ $\textup{cm}^{2}$ when the ions are near room temperature, resulting in $r=1.9 \times 10^3$ $\textup{s}^{-1}$. The electron-collisional excitation rate from the metastable state to the upper state can be estimated by considering electron-ion inelastic collisions \cite{curry_measurement_1995}
\begin{equation}
\label{eq:excitation}
u=\frac{6.5\times 10^{-4}}{\Delta ET_{\textup{e}}^{1/2}}n_{\textup{e}}f \exp\left ( -\frac{\Delta E}{T_{\textup{e}}} \right ),
\end{equation}
where $f$ is the oscillator strength \cite{griem_principles_2005}, $\Delta E$ is the change in energy of the transition (in eV), $T_{\textup{e}}$ is the electron temperature (in eV), and $n_{\textup{e}}$ is the electron density (in $\textup{cm}^{-3}$). Since this formula only provides an order of magnitude estimate of the collisional excitation rate, Ref.~\onlinecite{curry_measurement_1995} also provides a more appropriate rate $u \approx 1 \times 10^4$ $\textup{s}^{-1}$ based on the experiment. The sum of the theoretical quench rate and collisional excitation rate $r+\gamma  u$ can then be estimated as $1 \times 10^4$ $\textup{s}^{-1}$, which is consistent with our experimental value within errors. 

\section{Summary}
\label{sec:summary}

In this paper, we present the first experimental study of the metastable lifetime effects using wave-particle interaction and LIF in a multidipole plasma. The first numerical study of the hole-burning effect on the metastable distribution in the presence of an electrostatic wave is also reported. We demonstrate that the metastable lifetime and relative fraction of metastables produced from pre-existing ions can be determined though fitting the experiment results with the theory. The experimental finding verifies the previous simulation results that LIF performed on metastables produced directly from neutral atoms can only be used to infer the velocity distribution of ground-state ions if the ion dynamics is in a time shorter than the metastable lifetime. In the case where the metastable lifetime effects are inevitable, e.g., using LIF to measure the ion acoustic wave amplitude under 10 kHz in our experiment, we show that these systematic errors can be corrected with the theory addressed in the paper. Lastly, LIF measurements of $f_1$ and $f_2$ provide a new method to determine the absolute electric field without disturbing the plasma. This technique can be employed to calibrate other electric field measurement tools, such as the double-tip electric field probe used in our experiment.

\begin{acknowledgments}

This work was supported by the U.S. Department of Energy under Grant No. DE-SC0016473.

\end{acknowledgments}

\bibliography{refs}

\begin{thebibliography}{26}%
\makeatletter
\providecommand \@ifxundefined [1]{%
 \@ifx{#1\undefined}
}%
\providecommand \@ifnum [1]{%
 \ifnum #1\expandafter \@firstoftwo
 \else \expandafter \@secondoftwo
 \fi
}%
\providecommand \@ifx [1]{%
 \ifx #1\expandafter \@firstoftwo
 \else \expandafter \@secondoftwo
 \fi
}%
\providecommand \natexlab [1]{#1}%
\providecommand \enquote  [1]{``#1''}%
\providecommand \bibnamefont  [1]{#1}%
\providecommand \bibfnamefont [1]{#1}%
\providecommand \citenamefont [1]{#1}%
\providecommand \href@noop [0]{\@secondoftwo}%
\providecommand \href [0]{\begingroup \@sanitize@url \@href}%
\providecommand \@href[1]{\@@startlink{#1}\@@href}%
\providecommand \@@href[1]{\endgroup#1\@@endlink}%
\providecommand \@sanitize@url [0]{\catcode `\\12\catcode `\$12\catcode
  `\&12\catcode `\#12\catcode `\^12\catcode `\_12\catcode `\%12\relax}%
\providecommand \@@startlink[1]{}%
\providecommand \@@endlink[0]{}%
\providecommand \url  [0]{\begingroup\@sanitize@url \@url }%
\providecommand \@url [1]{\endgroup\@href {#1}{\urlprefix }}%
\providecommand \urlprefix  [0]{URL }%
\providecommand \Eprint [0]{\href }%
\providecommand \doibase [0]{http://dx.doi.org/}%
\providecommand \selectlanguage [0]{\@gobble}%
\providecommand \bibinfo  [0]{\@secondoftwo}%
\providecommand \bibfield  [0]{\@secondoftwo}%
\providecommand \translation [1]{[#1]}%
\providecommand \BibitemOpen [0]{}%
\providecommand \bibitemStop [0]{}%
\providecommand \bibitemNoStop [0]{.\EOS\space}%
\providecommand \EOS [0]{\spacefactor3000\relax}%
\providecommand \BibitemShut  [1]{\csname bibitem#1\endcsname}%
\let\auto@bib@innerbib\@empty
\bibitem [{\citenamefont {Hollmann}, \citenamefont {Anderegg},\ and\
  \citenamefont {Driscoll}(1999)}]{hollmann_measurement_1999}%
  \BibitemOpen
  \bibfield  {author} {\bibinfo {author} {\bibfnamefont {E.~M.}\ \bibnamefont
  {Hollmann}}, \bibinfo {author} {\bibfnamefont {F.}~\bibnamefont {Anderegg}},
  \ and\ \bibinfo {author} {\bibfnamefont {C.~F.}\ \bibnamefont {Driscoll}},\
  }\href {\doibase 10.1103/PhysRevLett.82.4839} {\bibfield  {journal} {\bibinfo
   {journal} {Phys. Rev. Lett.}\ }\textbf {\bibinfo {volume} {82}},\ \bibinfo
  {pages} {4839} (\bibinfo {year} {1999})}\BibitemShut {NoStop}%
\bibitem [{\citenamefont {Bowles}, \citenamefont {McWilliams},\ and\
  \citenamefont {Rynn}(1992)}]{bowles_direct_1992}%
  \BibitemOpen
  \bibfield  {author} {\bibinfo {author} {\bibfnamefont {J.}~\bibnamefont
  {Bowles}}, \bibinfo {author} {\bibfnamefont {R.}~\bibnamefont {McWilliams}},
  \ and\ \bibinfo {author} {\bibfnamefont {N.}~\bibnamefont {Rynn}},\ }\href
  {\doibase 10.1103/PhysRevLett.68.1144} {\bibfield  {journal} {\bibinfo
  {journal} {Phys. Rev. Lett.}\ }\textbf {\bibinfo {volume} {68}},\ \bibinfo
  {pages} {1144} (\bibinfo {year} {1992})}\BibitemShut {NoStop}%
\bibitem [{\citenamefont {Berumen}\ and\ \citenamefont
  {Skiff}(2018)}]{berumen_analysis_2018}%
  \BibitemOpen
  \bibfield  {author} {\bibinfo {author} {\bibfnamefont {J.}~\bibnamefont
  {Berumen}}\ and\ \bibinfo {author} {\bibfnamefont {F.}~\bibnamefont
  {Skiff}},\ }\href {\doibase 10.1063/1.5058805} {\bibfield  {journal}
  {\bibinfo  {journal} {Physics of Plasmas}\ }\textbf {\bibinfo {volume}
  {25}},\ \bibinfo {pages} {122102} (\bibinfo {year} {2018})}\BibitemShut
  {NoStop}%
\bibitem [{\citenamefont {Anderegg}\ \emph {et~al.}(1986)\citenamefont
  {Anderegg}, \citenamefont {Stern}, \citenamefont {Skiff}, \citenamefont
  {Hammel}, \citenamefont {Tran}, \citenamefont {Paris},\ and\ \citenamefont
  {Kohler}}]{anderegg_ion_1986}%
  \BibitemOpen
  \bibfield  {author} {\bibinfo {author} {\bibfnamefont {F.}~\bibnamefont
  {Anderegg}}, \bibinfo {author} {\bibfnamefont {R.~A.}\ \bibnamefont {Stern}},
  \bibinfo {author} {\bibfnamefont {F.}~\bibnamefont {Skiff}}, \bibinfo
  {author} {\bibfnamefont {B.~A.}\ \bibnamefont {Hammel}}, \bibinfo {author}
  {\bibfnamefont {M.~Q.}\ \bibnamefont {Tran}}, \bibinfo {author}
  {\bibfnamefont {P.~J.}\ \bibnamefont {Paris}}, \ and\ \bibinfo {author}
  {\bibfnamefont {P.}~\bibnamefont {Kohler}},\ }\href {\doibase
  10.1103/PhysRevLett.57.329} {\bibfield  {journal} {\bibinfo  {journal} {Phys.
  Rev. Lett.}\ }\textbf {\bibinfo {volume} {57}},\ \bibinfo {pages} {329}
  (\bibinfo {year} {1986})}\BibitemShut {NoStop}%
\bibitem [{\citenamefont {McChesney}, \citenamefont {Stern},\ and\
  \citenamefont {Bellan}(1987)}]{mcchesney_observation_1987}%
  \BibitemOpen
  \bibfield  {author} {\bibinfo {author} {\bibfnamefont {J.~M.}\ \bibnamefont
  {McChesney}}, \bibinfo {author} {\bibfnamefont {R.~A.}\ \bibnamefont
  {Stern}}, \ and\ \bibinfo {author} {\bibfnamefont {P.~M.}\ \bibnamefont
  {Bellan}},\ }\href {\doibase 10.1103/PhysRevLett.59.1436} {\bibfield
  {journal} {\bibinfo  {journal} {Phys. Rev. Lett.}\ }\textbf {\bibinfo
  {volume} {59}},\ \bibinfo {pages} {1436} (\bibinfo {year}
  {1987})}\BibitemShut {NoStop}%
\bibitem [{\citenamefont {Skiff}\ and\ \citenamefont
  {Anderegg}(1987)}]{skiff_direct_1987}%
  \BibitemOpen
  \bibfield  {author} {\bibinfo {author} {\bibfnamefont {F.}~\bibnamefont
  {Skiff}}\ and\ \bibinfo {author} {\bibfnamefont {F.}~\bibnamefont
  {Anderegg}},\ }\href {\doibase 10.1103/PhysRevLett.59.896} {\bibfield
  {journal} {\bibinfo  {journal} {Phys. Rev. Lett.}\ }\textbf {\bibinfo
  {volume} {59}},\ \bibinfo {pages} {896} (\bibinfo {year} {1987})}\BibitemShut
  {NoStop}%
\bibitem [{\citenamefont {Severn}\ \emph {et~al.}(2003)\citenamefont {Severn},
  \citenamefont {Wang}, \citenamefont {Ko},\ and\ \citenamefont
  {Hershkowitz}}]{severn_experimental_2003}%
  \BibitemOpen
  \bibfield  {author} {\bibinfo {author} {\bibfnamefont {G.~D.}\ \bibnamefont
  {Severn}}, \bibinfo {author} {\bibfnamefont {X.}~\bibnamefont {Wang}},
  \bibinfo {author} {\bibfnamefont {E.}~\bibnamefont {Ko}}, \ and\ \bibinfo
  {author} {\bibfnamefont {N.}~\bibnamefont {Hershkowitz}},\ }\href {\doibase
  10.1103/PhysRevLett.90.145001} {\bibfield  {journal} {\bibinfo  {journal}
  {Phys. Rev. Lett.}\ }\textbf {\bibinfo {volume} {90}},\ \bibinfo {pages}
  {145001} (\bibinfo {year} {2003})}\BibitemShut {NoStop}%
\bibitem [{\citenamefont {Booth}\ \emph {et~al.}(1989)\citenamefont {Booth},
  \citenamefont {Hancock}, \citenamefont {Perry},\ and\ \citenamefont
  {Toogood}}]{booth_spatially_1989}%
  \BibitemOpen
  \bibfield  {author} {\bibinfo {author} {\bibfnamefont {J.~P.}\ \bibnamefont
  {Booth}}, \bibinfo {author} {\bibfnamefont {G.}~\bibnamefont {Hancock}},
  \bibinfo {author} {\bibfnamefont {N.~D.}\ \bibnamefont {Perry}}, \ and\
  \bibinfo {author} {\bibfnamefont {M.~J.}\ \bibnamefont {Toogood}},\ }\href
  {\doibase 10.1063/1.343712} {\bibfield  {journal} {\bibinfo  {journal}
  {Journal of Applied Physics}\ }\textbf {\bibinfo {volume} {66}},\ \bibinfo
  {pages} {5251} (\bibinfo {year} {1989})}\BibitemShut {NoStop}%
\bibitem [{\citenamefont {McWilliams}\ \emph {et~al.}(2007)\citenamefont
  {McWilliams}, \citenamefont {Booth}, \citenamefont {Hudson}, \citenamefont
  {Thomas},\ and\ \citenamefont {Zimmerman}}]{mcwilliams_laser-induced_2007}%
  \BibitemOpen
  \bibfield  {author} {\bibinfo {author} {\bibfnamefont {R.}~\bibnamefont
  {McWilliams}}, \bibinfo {author} {\bibfnamefont {J.~P.}\ \bibnamefont
  {Booth}}, \bibinfo {author} {\bibfnamefont {E.~A.}\ \bibnamefont {Hudson}},
  \bibinfo {author} {\bibfnamefont {J.}~\bibnamefont {Thomas}}, \ and\ \bibinfo
  {author} {\bibfnamefont {D.}~\bibnamefont {Zimmerman}},\ }\href {\doibase
  10.1016/j.tsf.2006.10.027} {\bibfield  {journal} {\bibinfo  {journal} {Thin
  Solid Films}\ }\bibinfo {series} {The {{Third International Symposium}} on
  {{Dry Process}} ({{DPS}} 2005)},\ \textbf {\bibinfo {volume} {515}},\
  \bibinfo {pages} {4860} (\bibinfo {year} {2007})}\BibitemShut {NoStop}%
\bibitem [{\citenamefont {W.~A.~Hargus}\ and\ \citenamefont
  {Cappelli}(2001)}]{w._a._hargus_laser-induced_2001}%
  \BibitemOpen
  \bibfield  {author} {\bibinfo {author} {\bibfnamefont {J.}~\bibnamefont
  {W.~A.~Hargus}}\ and\ \bibinfo {author} {\bibfnamefont {M.~A.}\ \bibnamefont
  {Cappelli}},\ }\href {\doibase 10.1007/s003400100589} {\bibfield  {journal}
  {\bibinfo  {journal} {Appl Phys B}\ }\textbf {\bibinfo {volume} {72}},\
  \bibinfo {pages} {961} (\bibinfo {year} {2001})}\BibitemShut {NoStop}%
\bibitem [{\citenamefont {Mazouffre}, \citenamefont {Gawron},\ and\
  \citenamefont {Sadeghi}(2009)}]{mazouffre_time-resolved_2009}%
  \BibitemOpen
  \bibfield  {author} {\bibinfo {author} {\bibfnamefont {S.}~\bibnamefont
  {Mazouffre}}, \bibinfo {author} {\bibfnamefont {D.}~\bibnamefont {Gawron}}, \
  and\ \bibinfo {author} {\bibfnamefont {N.}~\bibnamefont {Sadeghi}},\ }\href
  {\doibase 10.1063/1.3112704} {\bibfield  {journal} {\bibinfo  {journal}
  {Physics of Plasmas}\ }\textbf {\bibinfo {volume} {16}},\ \bibinfo {pages}
  {043504} (\bibinfo {year} {2009})}\BibitemShut {NoStop}%
\bibitem [{\citenamefont {Claire}\ \emph {et~al.}(2001)\citenamefont {Claire},
  \citenamefont {Dindelegan}, \citenamefont {Bachet},\ and\ \citenamefont
  {Skiff}}]{claire_nonlinear_2001}%
  \BibitemOpen
  \bibfield  {author} {\bibinfo {author} {\bibfnamefont {N.}~\bibnamefont
  {Claire}}, \bibinfo {author} {\bibfnamefont {M.}~\bibnamefont {Dindelegan}},
  \bibinfo {author} {\bibfnamefont {G.}~\bibnamefont {Bachet}}, \ and\ \bibinfo
  {author} {\bibfnamefont {F.}~\bibnamefont {Skiff}},\ }\href {\doibase
  10.1063/1.1419221} {\bibfield  {journal} {\bibinfo  {journal} {Review of
  Scientific Instruments}\ }\textbf {\bibinfo {volume} {72}},\ \bibinfo {pages}
  {4372} (\bibinfo {year} {2001})}\BibitemShut {NoStop}%
\bibitem [{\citenamefont {Chu}\ \emph {et~al.}(2017)\citenamefont {Chu},
  \citenamefont {Mattingly}, \citenamefont {Berumen}, \citenamefont {Hood},\
  and\ \citenamefont {Skiff}}]{chu_determining_2017}%
  \BibitemOpen
  \bibfield  {author} {\bibinfo {author} {\bibfnamefont {F.}~\bibnamefont
  {Chu}}, \bibinfo {author} {\bibfnamefont {S.~W.}\ \bibnamefont {Mattingly}},
  \bibinfo {author} {\bibfnamefont {J.}~\bibnamefont {Berumen}}, \bibinfo
  {author} {\bibfnamefont {R.}~\bibnamefont {Hood}}, \ and\ \bibinfo {author}
  {\bibfnamefont {F.}~\bibnamefont {Skiff}},\ }\href {\doibase
  10.1088/1748-0221/12/11/C11005} {\bibfield  {journal} {\bibinfo  {journal}
  {J. Instrum.}\ }\textbf {\bibinfo {volume} {12}},\ \bibinfo {pages} {C11005}
  (\bibinfo {year} {2017})}\BibitemShut {NoStop}%
\bibitem [{\citenamefont {Chu}\ and\ \citenamefont
  {Skiff}(2018)}]{chu_lagrangian_2018}%
  \BibitemOpen
  \bibfield  {author} {\bibinfo {author} {\bibfnamefont {F.}~\bibnamefont
  {Chu}}\ and\ \bibinfo {author} {\bibfnamefont {F.}~\bibnamefont {Skiff}},\
  }\href {\doibase 10.1063/1.5020088} {\bibfield  {journal} {\bibinfo
  {journal} {Physics of Plasmas}\ }\textbf {\bibinfo {volume} {25}},\ \bibinfo
  {pages} {013506} (\bibinfo {year} {2018})}\BibitemShut {NoStop}%
\bibitem [{\citenamefont {Sarfaty}, \citenamefont {Souza-Machado},\ and\
  \citenamefont {Skiff}(1996)}]{sarfaty_direct_1996}%
  \BibitemOpen
  \bibfield  {author} {\bibinfo {author} {\bibfnamefont {M.}~\bibnamefont
  {Sarfaty}}, \bibinfo {author} {\bibfnamefont {S.~D.}\ \bibnamefont
  {Souza-Machado}}, \ and\ \bibinfo {author} {\bibfnamefont {F.}~\bibnamefont
  {Skiff}},\ }\href {\doibase 10.1063/1.871581} {\bibfield  {journal} {\bibinfo
   {journal} {Physics of Plasmas (1994-present)}\ }\textbf {\bibinfo {volume}
  {3}},\ \bibinfo {pages} {4316} (\bibinfo {year} {1996})}\BibitemShut
  {NoStop}%
\bibitem [{\citenamefont {McWilliams}\ and\ \citenamefont
  {Sheehan}(1986)}]{mcwilliams_experimental_1986}%
  \BibitemOpen
  \bibfield  {author} {\bibinfo {author} {\bibfnamefont {R.}~\bibnamefont
  {McWilliams}}\ and\ \bibinfo {author} {\bibfnamefont {D.}~\bibnamefont
  {Sheehan}},\ }\href {\doibase 10.1103/PhysRevLett.56.2485} {\bibfield
  {journal} {\bibinfo  {journal} {Phys. Rev. Lett.}\ }\textbf {\bibinfo
  {volume} {56}},\ \bibinfo {pages} {2485} (\bibinfo {year}
  {1986})}\BibitemShut {NoStop}%
\bibitem [{\citenamefont {Chu}\ and\ \citenamefont
  {Skiff}(2019)}]{chu_determining_2019}%
  \BibitemOpen
  \bibfield  {author} {\bibinfo {author} {\bibfnamefont {F.}~\bibnamefont
  {Chu}}\ and\ \bibinfo {author} {\bibfnamefont {F.}~\bibnamefont {Skiff}},\
  }\href {\doibase 10.1103/PhysRevLett.122.075001} {\bibfield  {journal}
  {\bibinfo  {journal} {Phys. Rev. Lett.}\ }\textbf {\bibinfo {volume} {122}},\
  \bibinfo {pages} {075001} (\bibinfo {year} {2019})}\BibitemShut {NoStop}%
\bibitem [{\citenamefont {Bennett}(1962)}]{bennett_hole_1962}%
  \BibitemOpen
  \bibfield  {author} {\bibinfo {author} {\bibfnamefont {W.~R.}\ \bibnamefont
  {Bennett}},\ }\href {\doibase 10.1103/PhysRev.126.580} {\bibfield  {journal}
  {\bibinfo  {journal} {Phys. Rev.}\ }\textbf {\bibinfo {volume} {126}},\
  \bibinfo {pages} {580} (\bibinfo {year} {1962})}\BibitemShut {NoStop}%
\bibitem [{\citenamefont {Hood}\ \emph {et~al.}(2016)\citenamefont {Hood},
  \citenamefont {Scheiner}, \citenamefont {Baalrud}, \citenamefont {Hopkins},
  \citenamefont {Barnat}, \citenamefont {Yee}, \citenamefont {Merlino},\ and\
  \citenamefont {Skiff}}]{hood_ion_2016}%
  \BibitemOpen
  \bibfield  {author} {\bibinfo {author} {\bibfnamefont {R.}~\bibnamefont
  {Hood}}, \bibinfo {author} {\bibfnamefont {B.}~\bibnamefont {Scheiner}},
  \bibinfo {author} {\bibfnamefont {S.~D.}\ \bibnamefont {Baalrud}}, \bibinfo
  {author} {\bibfnamefont {M.~M.}\ \bibnamefont {Hopkins}}, \bibinfo {author}
  {\bibfnamefont {E.~V.}\ \bibnamefont {Barnat}}, \bibinfo {author}
  {\bibfnamefont {B.~T.}\ \bibnamefont {Yee}}, \bibinfo {author} {\bibfnamefont
  {R.~L.}\ \bibnamefont {Merlino}}, \ and\ \bibinfo {author} {\bibfnamefont
  {F.}~\bibnamefont {Skiff}},\ }\href {\doibase 10.1063/1.4967870} {\bibfield
  {journal} {\bibinfo  {journal} {Physics of Plasmas}\ }\textbf {\bibinfo
  {volume} {23}},\ \bibinfo {pages} {113503} (\bibinfo {year}
  {2016})}\BibitemShut {NoStop}%
\bibitem [{\citenamefont {Severn}, \citenamefont {Edrich},\ and\ \citenamefont
  {McWilliams}(1998)}]{severn_argon_1998}%
  \BibitemOpen
  \bibfield  {author} {\bibinfo {author} {\bibfnamefont {G.~D.}\ \bibnamefont
  {Severn}}, \bibinfo {author} {\bibfnamefont {D.~A.}\ \bibnamefont {Edrich}},
  \ and\ \bibinfo {author} {\bibfnamefont {R.}~\bibnamefont {McWilliams}},\
  }\href {\doibase 10.1063/1.1148472} {\bibfield  {journal} {\bibinfo
  {journal} {Review of Scientific Instruments}\ }\textbf {\bibinfo {volume}
  {69}},\ \bibinfo {pages} {10} (\bibinfo {year} {1998})}\BibitemShut {NoStop}%
\bibitem [{\citenamefont {Mattingly}\ \emph {et~al.}(2013)\citenamefont
  {Mattingly}, \citenamefont {Berumen}, \citenamefont {Chu}, \citenamefont
  {Hood},\ and\ \citenamefont {Skiff}}]{mattingly_measurement_2013}%
  \BibitemOpen
  \bibfield  {author} {\bibinfo {author} {\bibfnamefont {S.~W.}\ \bibnamefont
  {Mattingly}}, \bibinfo {author} {\bibfnamefont {J.}~\bibnamefont {Berumen}},
  \bibinfo {author} {\bibfnamefont {F.}~\bibnamefont {Chu}}, \bibinfo {author}
  {\bibfnamefont {R.}~\bibnamefont {Hood}}, \ and\ \bibinfo {author}
  {\bibfnamefont {F.}~\bibnamefont {Skiff}},\ }\href {\doibase
  10.1088/1748-0221/8/11/C11015} {\bibfield  {journal} {\bibinfo  {journal} {J.
  Instrum.}\ }\textbf {\bibinfo {volume} {8}},\ \bibinfo {pages} {C11015}
  (\bibinfo {year} {2013})}\BibitemShut {NoStop}%
\bibitem [{\citenamefont {Goeckner}, \citenamefont {Goree},\ and\ \citenamefont
  {Sheridan}(1991)}]{goeckner_laserinduced_1991}%
  \BibitemOpen
  \bibfield  {author} {\bibinfo {author} {\bibfnamefont {M.~J.}\ \bibnamefont
  {Goeckner}}, \bibinfo {author} {\bibfnamefont {J.}~\bibnamefont {Goree}}, \
  and\ \bibinfo {author} {\bibfnamefont {T.~E.}\ \bibnamefont {Sheridan}},\
  }\href {\doibase 10.1063/1.859924} {\bibfield  {journal} {\bibinfo  {journal}
  {Physics of Fluids B: Plasma Physics (1989-1993)}\ }\textbf {\bibinfo
  {volume} {3}},\ \bibinfo {pages} {2913} (\bibinfo {year} {1991})}\BibitemShut
  {NoStop}%
\bibitem [{\citenamefont {Moore}(1963)}]{moore_physical_1963}%
  \BibitemOpen
  \bibfield  {author} {\bibinfo {author} {\bibfnamefont {W.~J.}\ \bibnamefont
  {Moore}},\ }\href@noop {} {\emph {\bibinfo {title} {Physical
  {{Chemistry}}}}}\ (\bibinfo {year} {1963})\BibitemShut {NoStop}%
\bibitem [{\citenamefont {Skiff}, \citenamefont {Bachet},\ and\ \citenamefont
  {Doveil}(2001)}]{skiff_ion_2001}%
  \BibitemOpen
  \bibfield  {author} {\bibinfo {author} {\bibfnamefont {F.}~\bibnamefont
  {Skiff}}, \bibinfo {author} {\bibfnamefont {G.}~\bibnamefont {Bachet}}, \
  and\ \bibinfo {author} {\bibfnamefont {F.}~\bibnamefont {Doveil}},\ }\href
  {\doibase 10.1063/1.1379044} {\bibfield  {journal} {\bibinfo  {journal}
  {Physics of Plasmas}\ }\textbf {\bibinfo {volume} {8}},\ \bibinfo {pages}
  {3139} (\bibinfo {year} {2001})}\BibitemShut {NoStop}%
\bibitem [{\citenamefont {Curry}\ \emph {et~al.}(1995)\citenamefont {Curry},
  \citenamefont {Skiff}, \citenamefont {Sarfaty},\ and\ \citenamefont
  {Good}}]{curry_measurement_1995}%
  \BibitemOpen
  \bibfield  {author} {\bibinfo {author} {\bibfnamefont {J.~J.}\ \bibnamefont
  {Curry}}, \bibinfo {author} {\bibfnamefont {F.}~\bibnamefont {Skiff}},
  \bibinfo {author} {\bibfnamefont {M.}~\bibnamefont {Sarfaty}}, \ and\
  \bibinfo {author} {\bibfnamefont {T.~N.}\ \bibnamefont {Good}},\ }\href
  {\doibase 10.1103/PhysRevLett.74.1767} {\bibfield  {journal} {\bibinfo
  {journal} {Phys. Rev. Lett.}\ }\textbf {\bibinfo {volume} {74}},\ \bibinfo
  {pages} {1767} (\bibinfo {year} {1995})}\BibitemShut {NoStop}%
\bibitem [{\citenamefont {Griem}(2005)}]{griem_principles_2005}%
  \BibitemOpen
  \bibfield  {author} {\bibinfo {author} {\bibfnamefont {H.~R.}\ \bibnamefont
  {Griem}},\ }\href@noop {} {\emph {\bibinfo {title} {Principles of {{Plasma
  Spectroscopy}}}}}\ (\bibinfo  {publisher} {{Cambridge University Press}},\
  \bibinfo {year} {2005})\BibitemShut {NoStop}%
\end{thebibliography}%

\end{document}